\documentclass[prb,aps,showpacs,twocolumn,unsortedaddress,longbibliography,superscriptaddress]{revtex4-1}%2
\usepackage{latexsym,amsmath,amssymb}
\usepackage{verbatim}
\usepackage{siunitx}
\usepackage[bookmarks=false,linkcolor=blue,urlcolor=blue,colorlinks,citecolor=blue]{hyperref}

\usepackage{graphicx}% Include figure files
\usepackage[dvipsnames]{xcolor}
\usepackage{color}
\usepackage{xspace}
\usepackage{soul}

\usepackage{dcolumn}% Align table columns on decimal point
\usepackage{bm}% bold math

\renewcommand{\selectlanguage}[1]{}

\begin{document}
\newcommand{\ovec}{\overrightarrow}

\title{Microwave-Tunable Diode Effect in Asymmetric SQUIDs with Topological Josephson Junctions}
\author{Joseph~J.~Cuozzo}
\thanks{jjcuozz@sandia.gov}
\author{Wei~Pan}
\affiliation{Materials Physics Department, Sandia National Laboratories, Livermore, CA 94551, USA.}
\author{Javad~Shabani}
\affiliation{Center for Quantum Information Physics, Department of Physics, New York University, NY 10003, USA.}
\author{Enrico~Rossi}
\affiliation{Department of Physics, William \& Mary, Williamsburg, VA 23187, USA}

\begin{abstract}
In superconducting systems in which inversion and time-reversal symmetry are simultaneously broken the critical current for positive and negative current bias can be different.
    For superconducting systems formed by Josephson junctions (JJs) this effect is termed Josephson diode effect.
    In this work, we study the Josephson diode effect for a superconducting quantum interference device (SQUID) formed by
    a topological JJ with a 4$\pi$-periodic current-phase relationship and a topologically trivial JJ.
    We show how the fractional Josephson effect manifests in the Josephson diode effect with the 
    application of a magnetic field and how tuning properties of the trivial SQUID arm can lead to diode polarity switching.
    We then investigate the AC response and show that the polarity of the diode effect can
    be tuned by varying the AC power and discuss differences between the AC diode effect of asymmetric SQUIDs
    with no topological JJ and SQUIDs in which one JJ is topological.
\end{abstract}

\maketitle

Recently there has been a great deal of activity investigating non-reciprocal effects and supercurrent rectification in 
superconductors~\cite{Silaev2014, Wakatsuki2017, Yasuda2019, Ando2020, Shin2021, Lyu2021, Daido2022, Yuan2022, Illic2022, Legg2022, Suri2022, He2022, Karabassov2022, Bauriedl2022, Strambini2022, Chahid2023} and Josephson junctions~\cite{Hu2007, Shi2015, Bocquillon2017, Pal2019, Misaki2021, Baumgartner2022_JPhys, Baumgartner2022_NatNano, Jeon2022, Halterman2022, Pal2022, Wu2022, Kokkeler2022, Zhang2022, Davydova2022, Illic2022_PRApplied, Trahms2023}. 
Conventional diodes, such as p-n junctions, have electrical resistance that depends on the direction of current and have numerous applications in computing, logic, and detection.
The superconducting diode effect (SDE) is characterized by a difference in forward and reverse critical currents $I_{+}$ and $I_{-}$ where the current range between $I_{+}$ and $I_{-}$ can be used to achieve supercurrent rectification. 
This non-reciprocal supercurrent develops due to simultaneous breaking of time-reversal and inversion symmetry~\cite{Onsager1931, Kubo1957, Rikken2001, Zhang2022}. 
Despite superconducting diodes having been discussed long ago~\cite{Krasnov1997, Touitou2004, Vodolazov2005, Papon2008, Carapella2010, Hu2007}, there has been a revival of interest, in part, due to signatures of finite-momentum Cooper pairing in helical superconductors~\cite{Davydova2022, Yuan2022, Pal2022} associated with the Josephson diode effect (JDE).
Superconducting diodes can also be used as passive on-chip gyrators, circulators, and memory in cryogenic applications~\cite{Golod2022}.

The fractional Josephson effect~\cite{Kitaev2003, Kwon2004} describes a 4$\pi$-periodic current-phase relationship in JJs originally associated with topological superconductivity.
Topological superconductivity has made important strides over the past decade since theoretical proposals to create topological superconductors for use in quantum computing have become feasible to realize~\cite{Nayak2008, fu_josephson_2009, Sau2010, Alicea2012, Pientka2017, Hell2017, aasen_milestones_2016}%. 
, although their discovery is still inconclusive~\cite{Lee2012_zb_anom, Kells2012, Cayao2015, Reeg2018, Penaranda2018, Vuik2019, Liu2019, Chen2019, Awoga2019, Woods2019, Prada2020, Valentini2021, Hess2021, Dartiailh2021a, microsoft2022}. 
Despite this, the fractional Josephson effect is well-documented in both topological~\cite{rokhinson_fractional_2012, Wiedenmann2016, Bocquillon2017, Yu2018, qu2022} and trivial JJs~\cite{Dartiailh2021}.
Furthermore, planar JJs are a suitable platform to realize a large JDE since both time-reversal and inversion symmetry can be readily and controllably broken~\cite{Kopasov2021, Legg2023}. 

In this article  
we study the DC and AC response of asymmetric SQUIDs~\cite{Fulton1972}.
Compared to previous studies we take into account effects due the SQUID's inductance, the presence of an AC bias,
and the role that a non-negligible fractional, $4\pi$, component of the current-phase-relation (CPR) for one of the JJ forming SQUIDs
has on the SQUID's diode effect.
We call a SQUID in which one JJ's CPR is $4\pi$, a 2$\pi$-4$\pi$ SQUID.
Recent experiments have shown that high-transparency wide JJs can also have a $4\pi$-periodic 
component of the current-phase relation~{\cite{Dartiailh2021}}. Our approach and results do not depend
on the origin of the $4\pi$-periodic component and therefore apply directly also to 
SQUIDs in which one JJ is wide and very transparent, as the one studied in Ref.~{\cite{Dartiailh2021}}.
First, we treat the problem with an analytic model that goes beyond the minimal models considered before~\cite{Fominov2022, Legg2023, Souto2022}.
We show that the DC response of $2\pi$-$4\pi$ SQUIDs exhibits 
the JDE and that the diode polarity is reversible with 
asymmetry in the normal resistance of the two SQUID arms.
We compare the JDE of a topological SQUID to a topologically-trivial one and find that, despite both SQUIDs showing comparable diode efficiencies, topological SQUIDs are of higher practical quality given they have a larger rectification current window $\Delta I_c$ coinciding with large diode efficiency
making them more robust to e.g. stray magnetic fields.
We also show the JDE can be switched and enhanced by an AC drive allowing for a microwave-controlled diode effect.
By including the inductance's effects we are able to properly characterize the ac response of the SQUID 
and show that the strength and sign of the diode effect depend on the ac power, an additional novel contribution toward
the understanding of the physics of asymmetric SQUIDs.
Lastly, we compare our analytic results with numerical simulations of the AC response of trivial asymmetric and 2$\pi$-4$\pi$ SQUIDs and find good agreement between the two approaches.

\begin{figure*}[ht!]
    \centering
    \includegraphics[width=1.99\columnwidth]{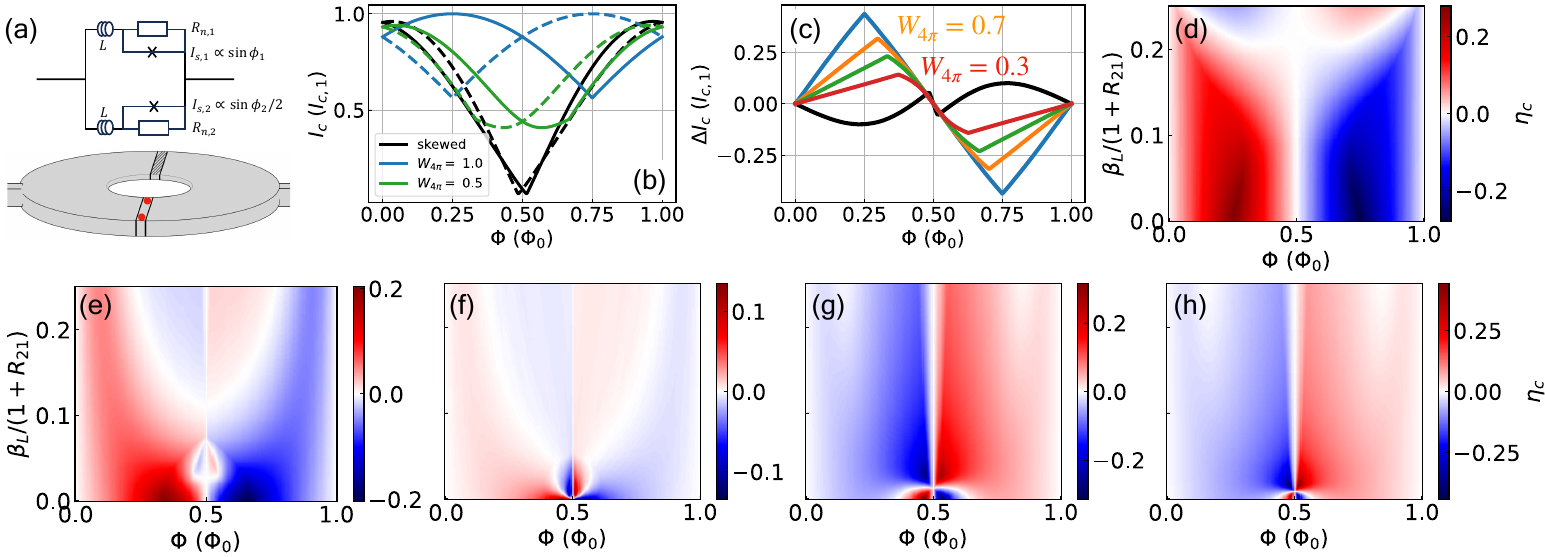} 
    \caption{ 
    (a) Circuit diagram of a 2$\pi$-4$\pi$ SQUID hosting Majorana zero modes in one arm. 
    (b) SQUID oscillations for $I_+$ (solid) and $I_-$ (dashed) with $\beta_L=0$. Skewed SQUID parameters are $a_1 = 1$ and $a_2 = 0.9 = 1 - c_2$ and
    (c) corresponding critical current difference for an asymmetric SQUID with $a_1 = 1$, $b_2 = W_{4\pi} = 1 - a_2$.
    $\eta_c$ dependence on $\Phi$ and $\frac{\beta_L}{1 + R_{21}}$ for $a_1 = 1$ 
    and (d) $W_{4\pi} = 1$, (e) $W_{4\pi} = 0.5$, (f) $W_{4\pi} = 0.1$, (g) $a_2 = 0.8$, $b_2 = 0.1 = c_2$, and (h) $a_2 = 0.9 = 1 - c_2$ (skewed SQUID).
    }
    \label{fig:dc_resp}
    \end{figure*}

    To model the dynamics of the JJs we use the resistively-shunted junction (RSJ) model:    
    $
        I_{B} = \frac{V_{J}}{R_n} + I_{s}
    $,
    where a current bias $I_B$ across a JJ is split into a resistive channel associated with quasiparticle current with normal resistance $R_n$ and a supercurrent channel $I_s$. Here we ignore charging effects associated with a capacitive channel. It is known that the Coulomb energy $E_C$ can compete with the Josephson energy $E_J$ in a 2$\pi$-4$\pi$ SQUID and lead to a gap in the mid-gap spectrum \cite{VanHeck2011} associated with quantum phase slips, reducing the 4$\pi$-periodicity to 2$\pi$. Here we assume $E_J > E_C$ for both SQUID arms, corresponding to wide topological JJs~\cite{Pientka2017, hell_two-dimensional_2017}. 

We can describe the fluxoid quantization condition with 
s-wave superconducting electrodes for the SQUID shown in Fig.~\ref{fig:dc_resp}(a). 
If the superconducting electrodes are thicker than the London penetration depth and the arms 
have equal inductance then we have the following current conservation and flux quantization conditions:
    $
    I_{B} = I_1 + I_2,
    $ $
    \phi_2 - \phi_1 = \frac{2\pi}{\Phi_0} \Phi_{tot}(\mathrm{mod}~2\pi)
    $
    where
    $
    \Phi_{tot} = L (I_1 - I_2) + \Phi
    $
     and 
    $
    I_k = \frac{V_{J,k}}{R_{n,k}} + I_{s,k},$ $k=1,2
    $.
    Here $I_1$ and $I_2$ are the currents in each of the SQUID arms, $\phi_1$ and $\phi_2$ are the gauge-invariant phase differences across each of the SQUID arms, $\Phi$ is the total external magnetic flux through the SQUID, $L$ is the inductance associated with the screening flux, $\Phi_0$ denotes the superconducting magnetic flux quantum $h/2e$, and $V_{J,k}$ and $I_{s,k}$ are the potential difference and the supercurrent of the $k^{th}$ arm, respectively. 
    In this work, we define an asymmetric SQUID as a SQUID with at least one of the following conditions: $I_{s,1}(\phi) \ne I_{s,2}(\phi)$, $I_{c,1} \ne I_{c,2}$, or $R_{n,1} \ne R_{n,2}$.
Using the Josephson relation $V_{J,k} = (\hbar /2e) \dot{\phi}_k$, we can combine these equations 
and solve for two coupled differential equations in terms of the 
average phase $\phi_A = (\phi_1 + \phi_2)/2$, and phase difference is $\Psi = (\phi_2 - \phi_1)/2\pi$
\begin{align}
        &\frac{d\phi_A}{d\tau} = \frac{1 + R_{21}}{4}i_B - \frac{i_{s,1}+\Delta_{21} i_{s,2}}{2} 
        + \frac{1-R_{21}}{4 \beta_L}\left(\Psi - \hat{\Phi} \right) 
        \label{eq:coupledODEs1}
\end{align}
\begin{align}
       &\frac{d\Psi}{d\tau} = \frac{R_{21} - 1}{4\pi}i_B + \frac{i_{s,1} - \Delta_{21} i_{s,2}}{2\pi} 
       - \frac{1+ R_{21}}{4 \pi \beta_L}\left(\Psi - \hat{\Phi} \right),
    \label{eq:coupledODEs}
\end{align}
$\tau = (2\pi I_{c,1}R_{n,1}/\Phi_0)t$ is a dimensionless time, 
$R_{21} = R_{n,2}/R_{n,1}$, 
$\Delta_{21} = R_{21}I_{c,2} / I_{c,1}$, 
$\beta_L = I_{c,1} L / \Phi_0$, 
$i_{s,k} = I_{s,k}/I_{c,1}$, and 
$\hat{\Phi} = \Phi/\Phi_0$. 
    
Evidence for non-sinusoidal terms contributing to a skewed CPR 
have been observed in past experiments \cite{Wiedenmann2016, Snyder2018, Lee2015, Panghotra2020, Dartiailh2021,qu2022}. To account for both the presence of skewed and topological CPRs, we assume a CPR with $\pi$-, $2\pi$-, and $4\pi$-periodic channels 
\begin{align}
    i_{s,1}(\phi_1) = &  ~a_1 \sin(\phi_1) + b_1 \sin\left(\frac{\phi_1}{2}\right) + c_1 \sin(2 \phi_1) \\
    \Delta_{21} i_{s,2}(\phi_2) & =  a_2 \sin(\phi_2) + b_2 \sin\left(\frac{\phi_2}{2}\right)  + c_2 \sin(2 \phi_2).
\end{align}
The $2\pi$ periodic term of the current phase relation is standard for a JJ, 
the $4\pi$-periodic contribution is present either from the topological character of the JJ or
from Landau-Zener transitions in high-transparency JJs, and the 
$\pi$-periodic term is the leading term that needs to be included
to describe JJs with good transparency.
For a ballistic short junction with a mode with transmission $\tau$, the CPR is described by
$
    I_s(\phi) \propto \sin\phi /\sqrt{1 - \tau \sin^2(\phi/2)},
$
where $0 \le \tau \le 1$ and $\phi$ is the phase across the junction. A Fourier expansion of the CPR to the second harmonic gives
$
    I_s(\phi) \approx I_1 \sin(\phi) + I_2 \sin(2 \phi),
$
where $I_1 > I_2$ and $I_2/I_1$ depends on $\tau$. For realistic values of $\tau$, we have $I_2 / I_1 \le 0.1$.
With this in mind, we constrain the amplitude $c_i \le 0.1$ in our calculations.
We assume $a_1 + b_1 + c_1 = 1$ and $a_2 + b_2 + c_2 = \Delta_{21}$ throughout the paper for simplicity. Furthermore, we assume $\Delta_{21} = 1$ throughout the 
article 
which implies the gaps of the junctions in the SQUID are the same. Following previous work \cite{Romeo2005}, we can reduce the SQUID dynamical equations to a single equation of motion by considering $\beta_L, ~\vert 1 - R_{21} \vert \ll 1$. Retaining terms linear in $\beta_L$, the SQUID dynamics are determined by the average phase $\phi_A$
\begin{equation}
    \frac{d\phi_A}{d\tau} = \frac{i_B}{2} - \tilde{i}_s(\phi_A) + \frac{\pi \beta_L (c_1 - c_2)^2}{2 (1+R_{21})} \sin(4\pi \hat{\Phi}),
    \label{eq:dPhiA}
\end{equation}
where
\begin{align}
    \tilde{i}_{s}(\phi_A) = & \sum_{m=1}^{6} \left[ x_m \sin\left(m\frac{\phi_A}{2}\right) + y_m \cos\left(m \frac{\phi_A}{2} \right) \right] \nonumber \\
    & + x_8 \sin\left(4 \phi_A\right) + y_8 \cos\left(4 \phi_A \right),
    \label{eq:asymSQUID}
\end{align}
$x_m$ and $y_m$ are coefficients that depend on $\hat{\Phi}$, $a_i$, $b_i$, $c_i$, and 
$\beta_L/(1+R_{21})$~\footnote{$x_m$ and $y_m$ are defined explicitly in Appendix C.}. 
The diode efficiency $\eta_c \equiv  (I_+ - I_-)/(I_+ + I_-)$ is often used to characterize superconducting diodes where the critical current $I_{+}$ ($-I_-$) corresponds to positive (negative) current bias. 
In an asymmetric SQUID, the broken chiral symmetry is due to different properties in the two arms of the SQUID, and the broken time-reversal symmetry is due to a magnetic flux threading the SQUID ring. For instance, in the topologically trivial asymmetric SQUID in Ref.~{\cite{Fominov2022}}, the diode is only present if an anomalous supercurrent exists at zero phase bias. This anomalous current breaks the chiral symmetry of the SQUID.
We extract $I_{\pm}$ from Eq.~(\ref{eq:dPhiA}) where the last two terms describe an effective CPR. First, it is worth noting that the effect of screening enters the dynamics via the term $\beta_L/(1+R_{21})$ suggesting an increase of $R_{21}$ is similar to a decrease of $\beta_L$.
The presence of $R_{21}$ in the effective CPR of the SQUID can be traced back to SQUID inductance contribution to the total magnetic flux where the currents $I_1$ and $I_2$ are currents which include both the supercurrent channels \textit{and} normal channels.
Second, the last term in Eq.~(\ref{eq:dPhiA}) is independent of $\phi_A$ but odd in $\hat{\Phi}$. This term applies an overall shift in the CPR which suggests a bipartite form of the diode effect 
$I_+ - I_- = \Delta \tilde{i}_{s,c} + \frac{\pi \beta_L (c_1 - c_2)^2}{2 (1+R_{21})} \sin(4\pi \hat{\Phi})$,
where the former term 
$\Delta \tilde{i}_{s,c} = \mathrm{max}\left(\tilde{i}_s \right) + \mathrm{min}\left(\tilde{i}_s \right)$
is determined by Eq.~(\ref{eq:asymSQUID}) and the latter is $\phi_A$-independent and associated with the screening current of imbalanced $\pi$ channels. In general, a SQUID with asymmetric skewed CPRs can expect additional contributions to the screening current term, and such shifts to the CPR can contribute
to anomalous scenarios such as $\vert \eta_c \vert >1$. 

We start by considering two types of SQUIDs. The first is a 2$\pi$-4$\pi$ SQUID with 4$\pi$ supercurrent in the topological arm characterized by the parameter $W_{4\pi} = b_2 = 1 - a_2$. The second is a trivial asymmetric SQUID (skewed SQUID) with $a_1 = 1$ and $a_2 = 0.9 = 1 - c_2$ ($b_1 = 0 = b_2$).

The DC responses of the SQUIDs are shown in Fig.~\ref{fig:dc_resp}(b). We notice that $I_c$ is largest when $\Phi=\Phi_0 /4$ for the 2$\pi$-4$\pi$ SQUID.
To understand this, recall that for a trivial SQUID with sinusoidal CPR's, the currents are maximized at $\phi_{max} = \pi/2$ and the two arms of the SQUID can simultaneously have that phase $\phi_{max}$ if the magnetic flux is an integer multiple of the magnetic flux quantum. Now, for the 2$\pi$-4$\pi$ SQUID, if the trivial arm has $\phi_{max,2\pi} = \pi/2$ and the non-trivial arm has $\phi_{max,4\pi} = \pi$, then it follows from the same argument that the maximum should occur at $\Phi_{ext} = \Phi_0/4$~\cite{Veldhorst2012, Legg2023}.

In Fig.~\ref{fig:dc_resp}(c), we present the difference in critical currents $\Delta I_c=I_+ - I_-$ for the 2$\pi$-4$\pi$ SQUID and trivial asymmetric SQUID considered in Fig.~\ref{fig:dc_resp}(b). A clear Josephson diode effect develops at $\Phi \ne n\Phi_0 /2$ ($n \in \mathbb{Z}$). Note, $\Delta I_c$ of the 2$\pi$-4$\pi$ SQUID exceeds that of the trivial asymmetric SQUID until $W_{4\pi} < 0.3$. 

    \begin{figure}[htb!]
    \centering
    \includegraphics[width=\linewidth]{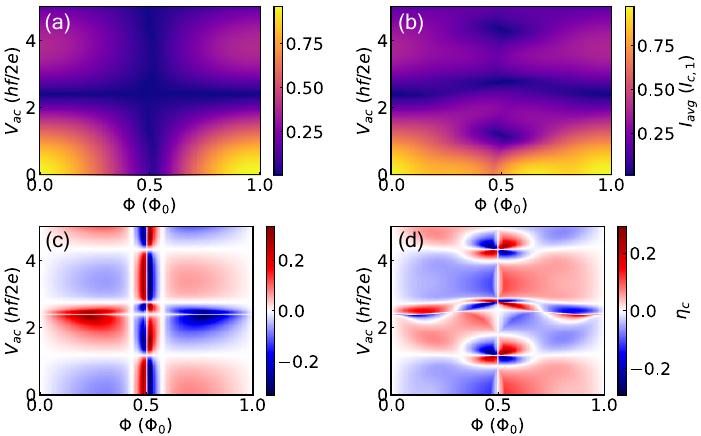}
    \caption{\label{fig:triv_squid}  
    AC power dependence of $I_c$ and $\eta_c$ for the skewed SQUID with (a,c) $\frac{\beta_L}{1+R_{21}} = 0$ and (b,d) $\frac{\beta_L}{1+R_{21}} = 0.125$.
    }
    \end{figure}

We present $\eta_c$ dependence on $\Phi$ and screening $\beta_L/(1 + R_{21})$ for 2$\pi$-4$\pi$ SQUIDs in Fig.~\ref{fig:dc_resp}(d-f). The diode efficiency of the 2$\pi$-4$\pi$ SQUID shown in Fig.~\ref{fig:dc_resp}(d) shows extrema for $\beta_L/(1+R_{21}) = 0$ and diode polarity switching for large screening. As $W_{4\pi}$ is decreased from unity (panels (e-f)), $\eta_c$ varies but the tunability of the diode polarity persists. 
Furthermore, as $W_{4\pi}$ decreases, the diode efficiency is generally smaller.

For a  SQUID nearly saturated with trivial supercurrent ($a_1 = 1$, $a_2 = 0.8$, and $b_2 = c_2 =0.1$), 
the regime of polarity switching with $\beta_L$ is pushed beyond our approximation of $\beta_L \ll 1$ (Fig.~\ref{fig:dc_resp}(g)) 
and closely resembles the trivial asymmetric SQUID DC response (Fig.~\ref{fig:dc_resp}(h)).
In the case of a trivial symmetric SQUID where $a_1 = 1 = a_2$, the diode efficiency $\eta_c = 0$ regardless 
of the value of $\Phi$ and $R_{21}$~\cite{Fominov2022}; this also holds for $\beta_L > 0$.
The source of the diode polarity switching with $\beta_L/(1+R_{21})$ is higher harmonic contributions to the CPR associated with the screening current ($\beta_L>0$). 
The inclusion of $\beta_L$ and $R_{21}$ is one of our main analytic results.
We also see that $\eta_c$ of the trivial asymmetric SQUID can be larger than $\eta_c$ of the 2$\pi$-4$\pi$ SQUID. 
The reason for this is that $\eta_c$ approaches unity when one of the critical currents approaches zero. 
Typically, this indicates an ideal diode, but if the non-zero critical current 
is also extremely small the practicality of such a diode is diminished since the current window for supercurrent rectification is also small.
Using $\vert \Delta I_c \vert$ as an additional quality-factor we find that the 2$\pi$-4$\pi$ SQUID diode outperforms the trivial asymmetric SQUID (See Appendix D and Fig.~\ref{fig:S2}).
The smallness of $I_c$ at half-flux is also the reason for the presence of strong variations, and polarity switchings, 
of $\eta_c$ when $\Phi/\Phi_0\approx 1/2$.
Such variations are physically uninteresting. 
In the remainder when discussing polarity switchings of
$\eta_c$ we refer to switchings at values of $\Phi/\Phi_0$ away from $1/2$.
We discuss $\eta_c$ in the remainder of the Letter for simplicity and comparison with the available literature, 
but we caution an over-emphasis on optimizing $\eta_c$ without consideration of the operational current range $\Delta I_c$.
Our results also suggest the control of the diode polarity with $R_{21}$ could be used as a signature of the fractional Josephson effect.

\begin{figure}[htb!]
 \centering
 \includegraphics[width=\linewidth]{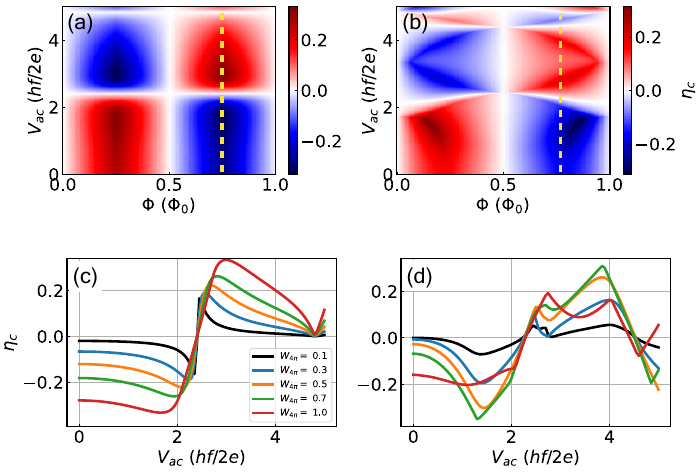}
 \caption{\label{fig:ac_diode}
 AC power dependence of $\eta_c$ for a 2$\pi$-4$\pi$ SQUID with (a) $\beta_L=0$ and (b) $\beta_L/(1+R_{21}) = 0.125$. 
 $\eta_c$ versus $V_{ac}$ at $\hat{\Phi}=3 /4$ for $a_1 = 1$ with (c) $\beta_L=0$ and (d) $\beta_L/(1 + R_{21}) = 0.125$.
 }
\end{figure}

To study the AC response of 
asymmetric SQUIDs we first consider the voltage-biased case, since
in this regime we can obtain analytical results.
Assuming
$V(t) = V_{dc} + V_{ac} \cos(2\pi f t)$, 
from the Josephson relation $\hbar d\phi_A/dt=2eV$, we obtain
$\phi_A(t) = \phi_0 + \omega_0 t + z\sin(2\pi f t)$ 
where $\phi_0$ is an arbitrary integration constant, 
$z=2eV_{ac} /(h f)$, and 
$\omega_0 = 2eV_{dc} /\hbar $.
Using Eqs.~(\ref{eq:dPhiA}-\ref{eq:asymSQUID}) we can obtain the $\bar I-V_{dc}$, with $\bar I$ being the time-averaged current, 
characteristic of the SQUID.
In the remainder we focus on the the behavior of the current when $V_{dc} = 0$.
   
Figures~\ref{fig:triv_squid}(a-b) show the SQUID critical current $I_{avg} \equiv I_+ + I_-$ as a function of $\Phi$ and $V_{ac}$ for $\beta_L/(1+R_{21})=0$ and $0.125$, respectively, for the skewed SQUID. 
In the absence of screening, $I_{avg}$ has a high degree of symmetry in ($\Phi$, $V_{ac}$) space defined by lines of $I_{avg} = 0$ at $\Phi = \Phi_0 / 2$ and $V_{ac}\sim 2.5~hf/2e$. With screening, lines of $I_{avg} = 0$ become broken and distorted. To see how this translates to the JDE, we present the corresponding diode efficiency in Fig.~\ref{fig:triv_squid}(c-d). We immediately notice the symmetry of $I_{avg}$ is preserved in $\eta_c$, particularly where $I_{avg} \sim 0$. In fact, $\eta_c$ has extrema near $I_{avg} \sim 0$ as a consequence of $I_{\pm} \rightarrow 0$ and $I_{\mp}>0$, as discussed earlier. 
We observe periodic diode polarity switching with increasing microwave power $V_{ac}$ for fixed $\Phi$.

    We can compare the AC response of the skewed SQUID of Fig.~\ref{fig:triv_squid} with a 2$\pi$-4$\pi$ SQUID shown in Fig.~\ref{fig:ac_diode}. Panel (a) and (b) show the AC response for $\beta_L/(1 + R_{21}) = 0$ and $0.125$, respectively. We notice the extrema of $\eta_c$ occur further away from $\Phi = \Phi_0/2$ compared to a trivial asymmetric SQUID, and the magnitude of $V_{ac}$ required to flip the diode polarity is generally larger than that of a trivial SQUID by a factor of two. The change in diode polarity can be attributed to the $J_0(z/2)$ Bessel function contribution to the gap, associated with the 4$\pi$ channel, which evolves with $z$ more slowly than the trivial Bessel dependence. Similar to a trivial asymmetric SQUID, a screening current distorts the symmetry of $\eta_c(\Phi,V_{ac})$. 
    
\begin{figure}[tb!]
    \centering
    \includegraphics[width=\linewidth]{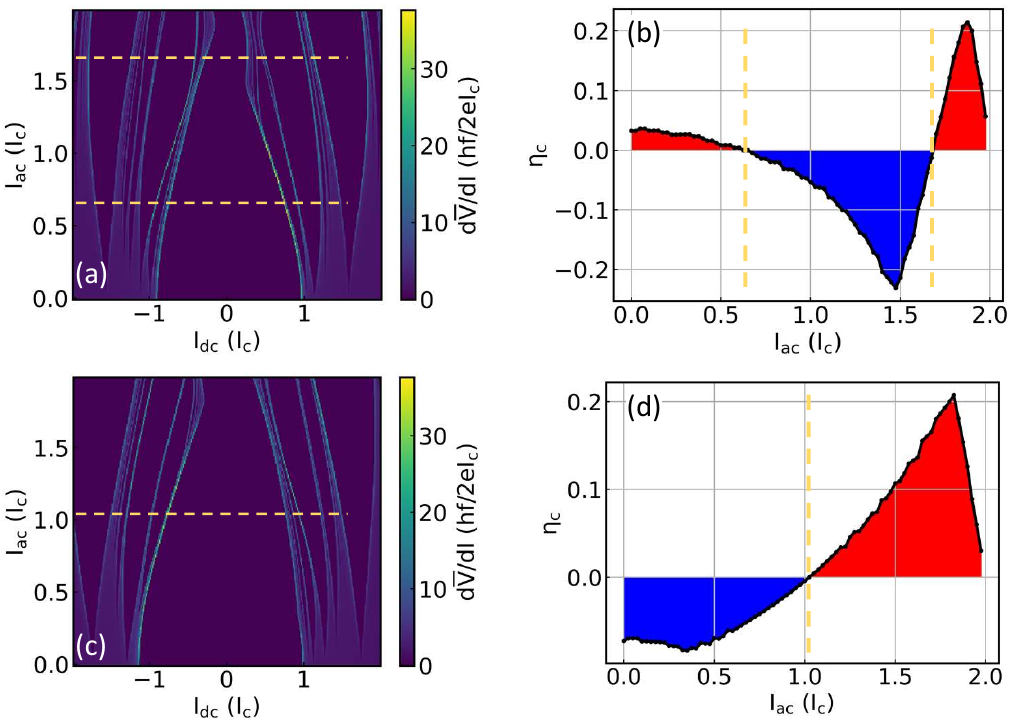}
    \caption{\label{fig:ibias}
    SQUID microwave response under current bias with $\hat{\Phi}= 3/4$ and $\beta_L = 1$ for
    (a-b) the skewed SQUID with $a_1 = 1$, $a_2 = 0.9 = 1-c_2$, $R_{21} = 2$ and
    (c-d) 2$\pi$-4$\pi$.
    Dashed lines indicate powers at which $\eta_c = 0$.
    }
    \end{figure}

    In Fig.~\ref{fig:ibias}, we consider the influence of microwave power in the experimentally-relevant current bias regime. We numerically solve the coupled system of non-linear differential equations described in Eq.~(\ref{eq:coupledODEs}) where we are not limited by the approximation $\beta_L,~ \vert 1-R_{21} \vert \ll 1$ used thus far. We consider a current bias $I_B = I_{dc} + I_{ac} \cos(2\pi ft)$ with a driving frequency $hf/\pi \Delta = 0.6$ where $\pi \Delta \equiv 2eI_c R_n$~\footnote{In terms of physical parameters, we can take I$_{c,1} = 0.5$ $\mu$A and R$_{n,1} = 85$ $\Omega$ ($E_{J,1} = 2e I_{c,1} R_{n,1} = 85$ $\mu$eV) at driving frequencies 2, 7 and 12 GHz.}. 
    
    Fig.~\ref{fig:ibias}(a) shows the power dependence of the $dV/dI$ characteristics for a trivial asymmetric SQUID ($\Phi = \Phi_0 / 4$) where the diode polarity gradually switches at high powers, as shown in panel (b). Dashed lines indicate a diode polarity switch. In agreement with Fig.~\ref{fig:triv_squid}(c-d), $\eta_c$ has a soft sign switch at low power before switching abruptly as the critical currents are nearly suppressed. Also in agreement with Fig.~\ref{fig:triv_squid}(c-d), $\eta_c$ has extrema as the critical currents are suppressed. Figure~\ref{fig:ibias}(c-d) presents the microwave response of a 2$\pi$-4$\pi$ SQUID. We note that the polarity of the 2$\pi$-4$\pi$ SQUID is opposite to that of the trivial asymmetric SQUID at zero AC power. $\eta_c$ has a weak enhancement in magnitude at lower
$I_{ac}$ before a gradual sign change at $I_{ac} = I_c$, which is at a higher power the first polarity switch of the asymmetric SQUID ($I_{ac} \sim 0.6 I_c$). Generally, the numerical results indicate good agreement with the analytic calculations.
    The $dV/dI$ characteristics are generally non-reciprocal, showing different Shapiro steps for positive and negative $I_{dc}$~\cite{Souto2022}. 
    
In this 
article, 
we studied the JDE in the DC and AC response of asymmetric SQUIDs, including the effects of inductance and asymmetries in $I_c$ and $R_n$. 
We showed that the inductance $\beta_L$ and the ratio $R_{21} = R_{n,2}/R_{n,1}$ can tune the diode efficiency of an asymmetric DC SQUID.
Such results may be applicable to recent experimental demonstrations of gate-tunable diode effects in asymmetric SQUIDs~{\cite{Ciaccia2023, Leblanc2023}}.
For SQUIDs with a 4$\pi$ junction, tuning $\beta_L$ and $R_{21}$ can cause a switching on the diode polarity.
We also showed a 2$\pi$-4$\pi$ SQUID has the opposite diode polarity of a trivial SQUID  over a wide range of $\beta_L/(1 + R_{21})$ and $\hat{\Phi}$.
We then discusseed how the Josephson diode polarity and efficiency of asymmetric SQUIDs 
can be controlled by 
microwave irradiation. 
We presented calculations of the AC response of asymmetric SQUIDs where the diode efficiency and polarity are controlled by the AC power.
The advantage of probing non-reciprocal transport in the AC response is that missing Shapiro steps indicative of a fractional Josephson effect have been observed experimentally~\cite{rokhinson_fractional_2012,Wiedenmann2016,Bocquillon2017,Dartiailh2021, Yu2018, qu2022}, suggesting the AC response of a 2$\pi$-4$\pi$ SQUID can readily be observed regardless of whether the 4$\pi$ junction is topological or not.
%    \\
\section*{Acknowledgments} 
    W.P, J.S., and E.R. acknowledge support from DOE, Grant No DE-SC0022245. The work at Sandia is supported by a LDRD project. Sandia National Laboratories is a multimission laboratory managed and operated by National Technology and Engineering Solutions of Sandia LLC, a wholly owned subsidiary of Honeywell International Inc.~for the U.S.~DOE's National Nuclear Security Administration under contract DE-NA0003525. This paper describes objective technical results and analysis. Any subjective views or opinions that might be expressed in the paper do not necessarily represent the views of the U.S. DOE or the United States Government.

\appendix

\section{2$\pi$-4$\pi$ SQUID dynamics}
    We start with the model for a semiclassical description of SQUID dynamics:
    \begin{eqnarray}
    I_{bias} &=& I_1 + I_2 \\
    \phi_2 - \phi_1 &=& \frac{2\pi}{\Phi_0} \Phi_{tot} \\
    \Phi_{tot} &=& L (I_1 - I_2) + \Phi \\
    I_i &=& \frac{V_{J,i}}{R} + I_{s,i} + C_{i} \frac{dV_{J,i}}{dt}
    \end{eqnarray}
    where $I_1$ and $I_2$ are the currents in each of the SQUID arms, $\phi_1$ and $\phi_2$ are the gauge-invariant phase differences across the JJ's in each of the SQUID arms, $\Phi$ is the total external magnetic flux through the SQUID, $L$ is the inductance associated with the screening flux, and $V_{J,i}$ and $I_{s,i}$ for $i=1,2$ are the potential difference across the $i^{th}$ JJ and the pair current in the $i^{th}$ JJ, respectively. We can consider the general RCSJ model for a SQUID device,
    \begin{subequations}
    \begin{align}
        \frac{d^2\phi_1}{d\tau^{\prime~2}} + \sigma \frac{d\phi_1}{d\tau^{\prime}} =& \frac{i_B}{2} - i_{s,~1}(\phi_1) \nonumber \\
    & + \frac{1}{4 \pi \beta_L} \left( \phi_2 - \phi_1 - 2\pi \hat{\Phi} \right)
    \end{align}
    \begin{align}
        C_{21} \frac{d^2\phi_2}{d\tau^{\prime~2}} + \frac{\sigma}{R_{21}}\frac{d\phi_2}{d\tau^{\prime}} = &\frac{i_B}{2} - I_{21} i_{s,~2}(\phi_2) \nonumber \\
    & - \frac{1}{4 \pi \beta_L} \left( \phi_2 - \phi_1 - 2\pi \hat{\Phi} \right)
    \end{align}
    \end{subequations}
    where $C_{21} = C_2 / C_1$, $I_{21} = I_{c,2}/ I_{c,1}$, $R_{21} = R_{n,2} / R_{n,1}$, $\sigma = \sqrt{\Phi_0/2\pi I_{c,1}R_{n,1}^2 C_1}$ and $\tau^{\prime} = \sqrt{2\pi I_{c,1}/\Phi_0C_1} t$. We will work in the overdamped regime for simplicity, but the extension is straightforward. Numerical calculations are generated by solving the system of coupled differential equations in the overdamped regime where capacitance is neglected.

\begin{figure*}[ht!]
\centering
\includegraphics[width=1.8\columnwidth]{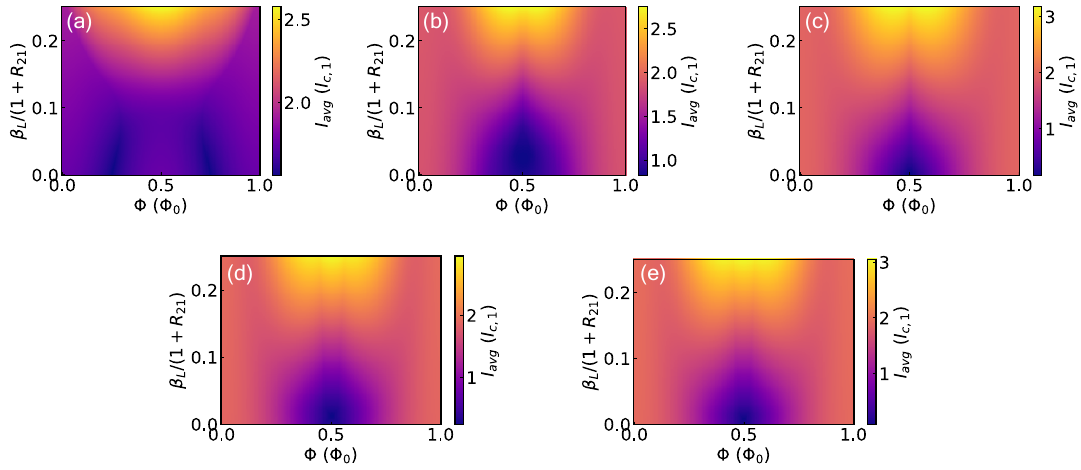} 
\caption{ 
$I_{avg}$ dependence on $\Phi$ and $\frac{\beta_L}{1 + R_{21}}$ for $a_1 = 1$ 
and (a-c) various values of $W_{4\pi}$, (e) $a_2 = 0.8$, $b_2 = 0.1 = c_2$, and (f) $a_2 = 0.9 = 1 - c_2$ (trivial SQUID).
}
\label{fig:S1}
\end{figure*}

    For a 2$\pi$-4$\pi$ SQUID, we consider the supercurrents $i_{s,~1} = \sin(\phi_1)$ and $i_{s,~2} = \sin(\phi_2/2)$ where $R_{21} = I_{21} =1$. We can reduce the SQUID dynamical equations to a single dynamical equation as a function of the average phase across the SQUID $\phi_A = (\phi_1 + \phi_2)/2$ by considering the inductance $\beta_L$ to be perturbatively small~\cite{Romeo2005}. The resulting dynamical equation is $i_B/2  = \frac{d\phi_A}{d\tau} + \tilde{i}_{s}(\phi_A, \hat{\Phi}_{ext})$ where $\tau \equiv (2\pi R I_{2\pi}/\Phi_0)t$ and,
    %\begin{widetext}
    \begin{align}
        \tilde{i}_s&(\phi_A, \hat{\Phi}) = \frac{1}{2} \sin\left( \frac{\phi_A+\pi \hat{\Phi}}{2} \right) + \frac{1}{2} \sin\left( \phi_A - \pi \hat{\Phi} \right) \\ \nonumber
        -~& \frac{\pi \beta_L}{8} \left[ 2 \sin\left( 2(\phi_A -\pi\hat{\Phi})\right) + \sin\left( \phi_A + \pi\hat{\Phi} \right)  \right] \\ \nonumber
        -~& \frac{\pi \beta_L}{8} \left[ \sin\left(\frac{\phi_A-3\pi \hat{\Phi}}{2} \right) - 3\sin\left(\frac{3\phi_A - \pi \hat{\Phi}}{2} \right) \right].
    \end{align}   
    %\end{widetext}
    
\subsection{DC Response}
    We find that the SQUID dc response to magnetic flux in the 2$\pi$-4$\pi$ SQUID is asymmetric: $I_{max}(\hat{\Phi}, I_{dc}) \ne I_{max}(-\hat{\Phi}, I_{dc}) \ne I_{max}(\hat{\Phi}, -I_{dc})$. The symmetry retained in the system is $I_{max}(\hat{\Phi}, I_{dc}) = I_{max}(-\hat{\Phi}, -I_{dc})$. 
    
    Besides this general asymmetry, we also notice that the maximum critical current does not manifest at $\Phi = 0$. Recall that for a trivial SQUID with sinusoidal CPR's, the currents are maximized at $\phi_{max} = \pi/2$ and the two arms of the SQUID can simultaneously have that phase $\phi_{max}$ if the magnetic flux is an integer multiple of the magnetic flux quantum:
    \begin{equation}
        \phi_2 - \phi_1 = \frac{2 \pi \Phi}{\Phi_0} \left( \mathrm{mod}~2\pi \right)
    \end{equation}
    Now, for the 2$\pi$-4$\pi$ SQUID, if the trivial arm (say, arm 1) has $\phi_{max,1} = \pi/2$ and the non-trivial arm has $\phi_{max,2} = \pi$, then it follows from the argument for the trivial SQUID that the maximum should occur at $\Phi = \Phi_0/4$.

\section{Symmetric SQUID with $\pi$-, $2\pi$-, \& $4\pi$-periodic Channels} 
In this section, we provide the general solution for a symmetric DC SQUID circuit model with negligible capacitance, weak inductance, and a supercurrent with $\pi$-, $2\pi$-, and $4\pi$-periodic channels. We write an effective description of the supercurrent channel with a skewed CPR and a topological contribution as,
\begin{equation}
    I_s = I_{4\pi} \sin(\phi/2) + I_{2\pi} \sin(\phi) + I_{\pi} \sin(2\phi).
    \label{pairchannel}
\end{equation}
Making use of the ac Josephson effect $\frac{d\phi}{dt} = \frac{2e}{\hbar}V$, we find
\begin{widetext}
\begin{align}
\frac{d\phi_1}{d\tau} + \sin(\phi_1) + \tilde{\beta} \sin(2\phi_1) + \alpha \sin(\phi_1/2) + \frac{\phi_1 - \phi_2}{4\pi \beta_L} & = \frac{1}{2} \left( i_B  - \frac{\hat{\Phi}}{\beta_L} \right)
\label{arm1} \\
\frac{d\phi_2}{d\tau} + \sin(\phi_2) + \tilde{\beta} \sin(2\phi_2) + \alpha \sin(\phi_2/2) - \frac{\phi_1 - \phi_2}{4\pi \beta_L} & = \frac{1}{2} \left( i_B  + \frac{\hat{\Phi}}{\beta_L} \right)
\label{arm2}
\end{align}
\end{widetext}
where $\tilde{\beta}\equiv I_{\pi}/I_{2\pi}$, 
$\alpha \equiv I_{4\pi}/I_{2\pi}$, 
$\beta_L \equiv LI_{2\pi}/\Phi_0$,\\
$i_B \equiv I_{bias}/I_{2\pi}$, 
$\hat{\Phi} \equiv \Phi
/\Phi_0$, and $\tau \equiv (2\pi R_{n,1} I_{2\pi}/\Phi_0)t$. Defining $\phi_A \equiv (\phi_1 + \phi_2)/2$ and $\Psi \equiv (\phi_2 - \phi_1)/2\pi$, we can consider the sum and difference of equations to find

\begin{widetext}
\begin{align}
    \frac{d\phi_A}{d\tau} + \sin(\phi_A)\cos(\pi \Psi) + \tilde{\beta} \sin(2\phi_A)\cos(2\pi \Psi) + \alpha \sin(\phi_A/2)\cos(\pi \Psi/2) & = \frac{i_B}{2}
    \label{Ibias} \\
    \pi \frac{d\Psi}{d\tau} + \frac{\Psi}{2 \beta_L} + \sin(\pi \Psi)\cos(\Phi_A) + \tilde{\beta} \sin(2\pi \Psi) \cos(2\phi_A) + \alpha \sin(\pi \Psi/2) \cos(\phi_A/2) & = \frac{\hat{\Phi}}{2 \beta_L}
\end{align}    
\end{widetext}
Assuming $\beta_L \ll 1$, we make the following ansatz:
\begin{equation}
    \Psi(\tau) = \hat{\Phi} + \beta_L \Psi_1(\tau) + \mathcal{O}(\beta_L^2)
\end{equation}

Substituting, we find the solution to lowest order in $\beta_L$ is
\begin{align}
    \Psi_1(\tau) &= -2 [ \alpha \sin(\pi \hat{\Phi}/2) \cos(\phi_A/2) \nonumber \\
    & + \tilde{\beta} \sin(2 \pi \hat{\Phi}) \cos(2\phi_A) + \sin(\pi \hat{\Phi}) \cos(\phi_A) ].
    \label{PsiSoln}
\end{align}
Now we can reduce the system of coupled equations into a single equation for $\phi_A$ and calculate the time-averaged current bias for an rf-driven junction. Substituting Eq. \ref{PsiSoln} into Eq. \ref{Ibias} and simplifying, we find
\begin{align}
    \frac{d\phi_A}{d\tau} & + a \sin(\phi_A) + b \sin(2\phi_A) + c\sin(3\phi_A) + d \sin(4\phi_A) \nonumber \\
    & + f \sin\left(\frac{\phi_A}{2}\right) + g \sin\left(\frac{3\phi_A}{2}\right) + h\sin\left(\frac{5\phi_A}{2}\right) = \frac{i_B}{2}
    \label{soln}
\end{align}
for the coefficients,
\begin{align}
    a & = x(1- \pi \beta_L \tilde{\beta} y^2) + \frac{\pi}{4} \alpha^2 \beta_L (1-x) \\
    b & = \tilde{\beta} + (\pi \beta_L - 2\tilde{\beta})y^2 \\
    c & = 6 \pi \beta_L \tilde{\beta} x y^2 \\
    d & = 2 \pi \beta_L \tilde{\beta}^2y^2 \\
    f & = \alpha (C_{\pi/2} + \frac{\pi}{2}\beta_L y S_{\pi/2}) \\
    g & = \frac{3\pi}{2} \alpha \beta_L y (1+2\tilde{\beta}x)S_{\pi/2} \\
    h & = 5 \pi \alpha \beta_L \tilde{\beta} x y S_{\pi/2}
\end{align}
where $S_{\pi/2} \equiv \sin(\pi \hat{\Phi}/2)$, $C_{\pi/2} \equiv \cos(\pi \hat{\Phi}/2)$, $x \equiv \cos(\pi \hat{\Phi})$, and $y \equiv \sin(\pi \hat{\Phi})$. Note that if $\hat{\Phi} = 0$, then only $a$, $b$ and $f$ are non-zero. The coefficients in Eq.~(\ref{soln}) have the following interpretations:
\begin{itemize}
    \item $a$: $2\pi$ channel of each arm, the \textit{interference} of the $4\pi$ channels of the arms, and \textit{interference} of the $2\pi$ and $\pi$ channels of the arms
    \item $b$: $\pi$ channel of each arm and the \textit{interference} of the $2\pi$ channels of the arms
    \item $c$: \textit{interference} of the $2\pi$ and $\pi$ channels of the arms
    \item $d$: \textit{interference} of the $\pi$ channels of the arms
    \item $f$: $4\pi$ channel of each arm
    \item $g$: \textit{interference} of $4\pi$ and $2\pi$ channels of the arms, and the \textit{interference} of $4\pi$ and $\pi$ channels of the arms
    \item $h$: \textit{interference} of $4\pi$ and $\pi$ channels of the arms
\end{itemize}

\subsection{Voltage-bias solution}
From here, we can consider a voltage bias
\begin{equation}
    V(\tau) = V_0 + V_1 \cos(\omega \tau)
\end{equation}
and make use of the ac Josephson effect 
\begin{equation*}
    \frac{d\phi_A}{dt} = \frac{2e}{\hbar}V
\end{equation*}
to solve for $\phi_A(\tau)$ and substitute into Eq. \ref{soln}. Then we can use the Jacobi-Anger expansion,
\begin{equation}
    e^{iz\sin(\theta)} = \sum_{n=-\infty}^{+\infty} J_{n}(z) e^{in\theta},
\end{equation}
where $J_n$ are $n^{th}$ order Bessel functions, to calculate the Shapiro spikes and each spike's width. 

\begin{figure*}[ht!]
\centering
\includegraphics[width=1.8\columnwidth]{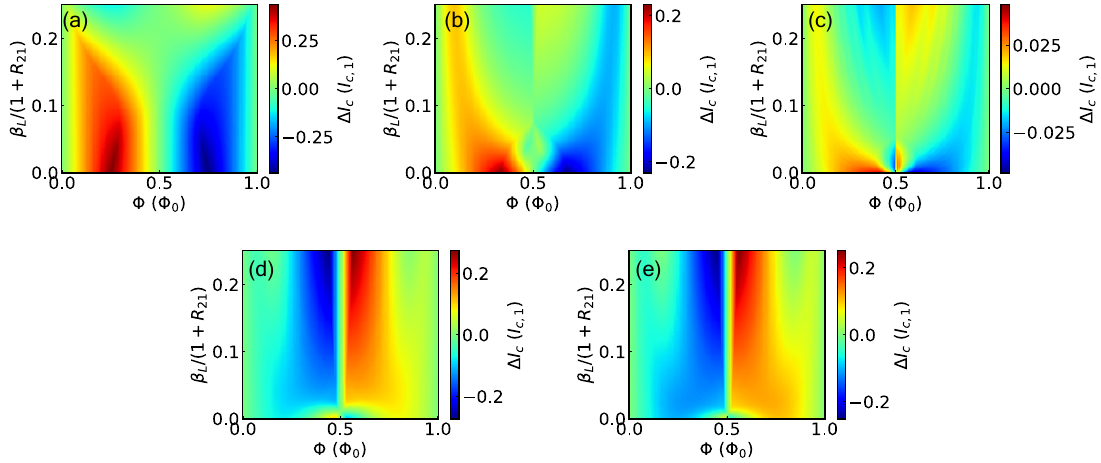} 
\caption{ 
$\Delta I_c$ dependence on $\Phi$ and $\frac{\beta_L}{1 + R_{21}}$ for $a_1 = 1$ 
and (a-c) various values of $W_{4\pi}$, (e) $a_2 = 0.8$, $b_2 = 0.1 = c_2$, and (f) $a_2 = 0.9 = 1 - c_2$ (trivial SQUID).
}
\label{fig:S2}
\end{figure*}

Now we will describe how to calculate Shapiro spike widths in terms of the time-averaged pair current $\overline{I}_s$ and specifically consider the $n=0$ spike. We start by integrating the ac Josephson effect from the end of the previous section. We can write (in dimensionless parameters),
\begin{equation}
    \phi_A(\tau) = \phi_0 + \omega_0 \tau + z\sin(\omega \tau)
\end{equation}
where $\phi_0$ is an arbitrary integration constant, $z = 2eV_{1}/ \hbar \omega$, and $\omega_0 = 2 e V_{0}/\hbar$. 
We then substitute into Eq. (\ref{soln}) to get $2d\phi_A/d\tau + I_{s} = i_B$ where
\begin{widetext}
\begin{align}
    I_s & = 2 \mathrm{Im}\{ \sum_{n=-\infty}^{+\infty} (-1)^n e^{-i n\omega \tau} [a e^{i(\phi_0 + \omega_0 \tau)} J_n(z) + b e^{2i(\phi_0 + \omega_0 \tau)} J_n(2z) + c e^{3i(\phi_0 + \omega_0 \tau)} J_n(3z) \nonumber \\
    &+ d e^{4i(\phi_0 + \omega_0 \tau)} J_n(4z) + f e^{\frac{1}{2}i(\phi_0 + \omega_0 \tau)} J_n(z/2) + g e^{\frac{3}{2}i(\phi_0 + \omega_0 \tau)} J_n(3z/2) + h e^{\frac{5}{2}i(\phi_0 + \omega_0 \tau)} J_n(5z/2)] \}
\end{align}
\end{widetext}

\section{Asymmetric SQUID dynamics}
    Now we assume a general CPR with $\pi$-, $2\pi$-, and $4\pi$-periodic channels,
    \begin{align}
        i_{s,1}(\phi_1) & = a_1 \sin(\phi_1) + b_1 \sin(\frac{\phi_1}{2}) + c_1 \sin(2 \phi_1) %\\
    \end{align}
    \begin{align}
        \Delta_{21} i_{s,2}(\phi_2) & = a_2 \sin(\phi_2) + b_2 \sin(\phi/2) + c_2 \sin(2 \phi_2)
    \end{align}
    where $\Delta_{21} = I_{c,2}R_{n,2}/I_{c,1}R_{n,1}$. If we assume $\beta_L,~\vert 1-R_{21}\vert \ll 1$ then we can reduce the system of 2 ODE's to a single ODE via perturbative ansatz similar to the ansatz made by de Luca:
    \begin{equation}
        \frac{d\phi_A}{d\tau} = \frac{i_B}{2} - \tilde{i}_s(\phi_A) + \frac{\pi \beta_L (c_1 - c_2)^2}{2 (1+R_{21})} S_{4}
    \end{equation}
    where
    \begin{widetext}
    \begin{align}
        \tilde{i}_s(\phi_A) & =  x_2 \sin(\phi_A) + x_4 \sin(2 \phi_A) + x_6 \sin(3 \phi_A) + x_8 \sin(4 \phi_A) + x_1 \sin(\phi_A/2) %\nonumber \\ 
        %& 
        + x_3 \sin(3\phi_A/2) + x_5 \sin(5\phi_A/2)  \nonumber \\
        &+ y_2 \cos(\phi_A) + y_4 \cos(2 \phi_A) + y_6 \cos(3 \phi_A) + y_8 \cos(4 \phi_A) %\nonumber \\  
        %& 
        + y_1 \cos(\phi_A/2) + y_3 \cos(3\phi_A/2) + y_5 \cos(5\phi_A/2).
    \end{align}
    \end{widetext}

    The coefficients of the effective supercurrent $\tilde{i}_s$ are ($C_n \equiv \cos(n\pi \hat{\Phi})$ and $S_n \equiv \sin(n\pi \hat{\Phi})$):
    \begin{widetext}
    \begin{align}
        x_2 & =  \frac{\pi \beta_L b_1 b_2}{2(1+R_{21})} + \left[ \frac{a_1 + a_2}{2} - \frac{\pi \beta_L}{1+R_{21}}\left( \frac{b_1^2}{2} - a_1c_1 + a_1c_2 + \frac{b_2^2}{4} - \frac{a_2c_1}{2} \right) \right] C_1 %\nonumber \\
        %& 
        - \frac{\pi \beta_L}{1+R_{21}}\left( a_2c_1 - \frac{a_2c_2}{2} + \frac{a_1c_1}{2} \right) C_3 \\
        y_2 & = \left[ \frac{a_2 - a_1}{2} - \frac{\pi \beta_L}{1+R_{21}}\left( a_1c_1 - \frac{a_1c_2}{2}+\frac{b_2^2}{4} - \frac{a_2c_1}{2} \right) \right] S_1 %\nonumber \\ 
        %& 
        - \frac{\pi \beta_L}{1+R_{21}}\left( \frac{a_2c_2}{2} - a_2c_1 +\frac{a_1c_1}{2} \right) S_3 \\
        x_4 & = \frac{\pi \beta_L a_1 a_2}{1+R_{21}} + \left( \frac{c_1+c_2}{2} - \frac{\pi \beta_L(a_1^2 + a_2^2)}{2(1+R_{21})} \right) C_2 \\
        y_4 & = \left( \frac{c_2-c_1}{2} - \frac{\pi \beta_L (a_2^2 - a_1^2)}{4(1+R_{21})} \right) S_2 \\
        x_6 & = -\frac{\pi \beta_L}{1+R_{21}}\left( \frac{a_2c_2}{2} - 2a_2c_1 - \frac{a_1c_1}{2} - a_1c_2 \right)C_1 %\nonumber \\
        %& 
        - \frac{\pi \beta_L}{1+R_{21}} \left( 2a_1c_1 + \frac{a_2c_1}{2} + a_2c_2 + \frac{a_1c_2}{2} \right) C_3 \\
        y_6 & = -\frac{\pi \beta_L}{1+R_{21}}\left( 2a_2c_1 - \frac{a_2c_2+a_1c_1}{2} - a_1c_2 \right)S_1 %\nonumber \\
        %& 
        - \frac{\pi\beta_L}{1+R_{21}}\left( -2a_1c_1 + \frac{a_2c_1}{2} + a_2c_2 + \frac{a_1c_2}{2} \right)S_3 \\
        \end{align}
        \begin{align}
        x_8 & = -\frac{\pi \beta_L}{1+R_{21}}\left( \frac{c_2^2-c_1^2}{2} - 2c_1c_2 \right) - \frac{\pi \beta_L}{1+R_{21}} \left( \frac{3c_1^2+c_2^2}{2} \right) C_4 \\
        y_8 & = -\frac{\pi \beta_L}{1+R_{21}}\left( \frac{-3c_1^2+2c_1c_2+c_2^2}{2} \right) S_4 \\
        x_1 & = \left(\frac{b_1+b_2}{2}+\frac{\pi \beta_L}{1+R_{21}} \frac{a_1b_1+a_2b_2}{4} \right) C_{1/2} - \frac{\pi \beta_L}{1+R_{21}}\left( \frac{a_1b_2+a_2b_1}{4} \right) C_{3/2} \\
        y_1 & = \left(\frac{b_2-b_1}{2}+\frac{\pi \beta_L}{1+R_{21}} \frac{a_2b_2-a_1b_1}{4} \right) S_{1/2} - \frac{\pi \beta_L}{1+R_{21}}\left( \frac{a_2b_1-a_1b_2}{4} \right) S_{3/2} \\
        x_3 & = \frac{\pi \beta_L}{1+R_{21}}[ \left( \frac{3a_1b_2 + 3a_2b_1}{4} \right) C_{1/2} - \left( \frac{3a_1b_1-5b_1c_1+2b_1c_2+3a_2b_2-2b_2c_1-b_2c_2}{4} \right) C_{3/2} \nonumber \\
        & - \left( \frac{-2b_2c_2 + 2b_1c_1 + b_1c_2+5b_2c_1}{4} \right) C_{5/2} ] \\
        y_3 & = -\frac{\pi \beta_L}{1+R_{21}} [ \left( \frac{3(a_1b_2 -a_2b_1)}{4} \right)S_{1/2} +\left( \frac{-3a_1b_1 + 5b_1c_1 -2b_1c_2 +3a_2b_2 -2b_2c_1 -b_2c_2}{4} \right)S_{3/2} \nonumber \\
        & + \left( \frac{2b_2c_2 + 2b_1c_1 + b_1c_2 - 5b_2c_1}{4} \right)S_{5/2} ] \\
        x_5 & = -\frac{\pi \beta_L}{1+R_{21}}\left[ \left( \frac{-7b_2c_1 + 2b_2c_2 - 2b_1c_1 - 3b_1c_2}{4} \right)C_{3/2} + \left( \frac{7b_1c_1 -2b_1c_2 + 2b_2c_1 + 3b_2c_2}{4} \right)C_{5/2} \right] \\
        y_5 & = -\frac{\pi \beta_L}{1+R_{21}}\left[ \left( \frac{7b_2c_1 - 2b_2c_2 - 2b_1c_1 - 3b_1c_2}{4} \right)S_{3/2} + \left( \frac{-7b_1c_1 + 2b_1c_2 + 2b_2c_1 + 3b_2c_2}{4} \right)S_{5/2} \right]
    \end{align}
    \end{widetext}
    
    These are complicated expressions, but we can gain insight about the effects of asymmetry on the dc and ac response of the SQUID. Firstly, we notice that the harmonics entering the effective supercurrent are the same as those in the symmetric case, except here we have both cosine and sine terms. For $\hat{\Phi} = 0$, we have $y_1 = ...=y_8 = 0$ so that only the sine terms contribute to the zero-field SQUID response. Interestingly, higher harmonics (e.g. $\sin(2 \phi_A)$) contribute to the effective supercurrent at zero-field as opposed to the symmetric case where higher harmonic contributions only affect the SQUID response at nonzero magnetic flux. 
    
\begin{figure}[ht!]
\centering
\includegraphics[width=0.95\columnwidth]{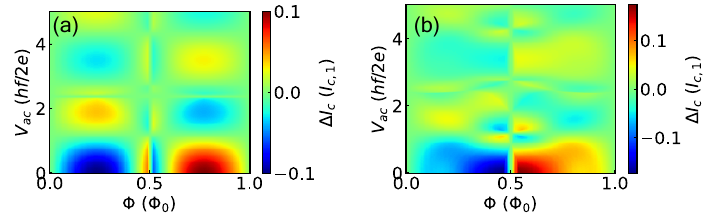} 
\caption{ 
AC power dependence of $\Delta I_c$ for trivial SQUID with (a) $\frac{\beta_L}{1+R_{21}} = 0$ and (b) $\frac{\beta_L}{1+R_{21}} = 0.125$.
}
\label{fig:S3}
\end{figure}

\begin{figure}[ht!]
\centering
\includegraphics[width=0.95\columnwidth]{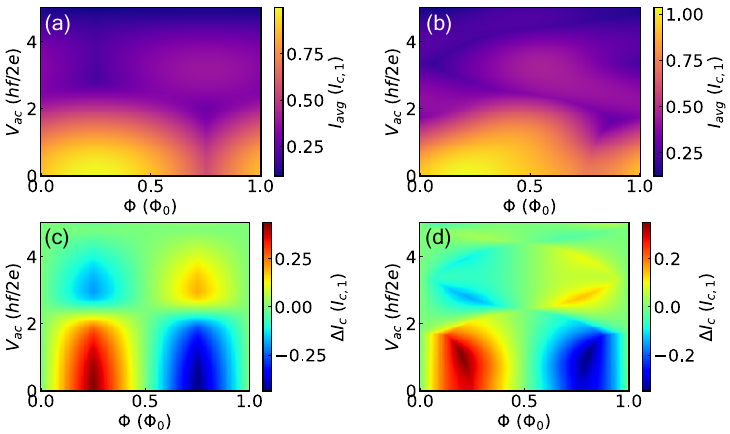} 
\caption{ 
AC power dependence of $I_{avg}$ and $\Delta I_c$ for 2$\pi$-4$\pi$ SQUID with (a) $\frac{\beta_L}{1+R_{21}} = 0$ and (b) $\frac{\beta_L}{1+R_{21}} = 0.125$.
}
\label{fig:S4}
\end{figure}

\subsection{Voltage-bias solution}

As before, we can consider a voltage bias
\begin{equation}
    V(\tau) = V_0 + V_1 \cos(\omega \tau)
\end{equation}
and make use of the ac Josephson effect 
\begin{equation}
    \frac{d\phi_A}{dt} = \frac{2e}{\hbar}V
\end{equation}
to solve for $\phi_A(\tau)$. Then we can use the Jacobi-Anger expansion,
\begin{equation}
    e^{iz\sin(\theta)} = \sum_{n=-\infty}^{+\infty} J_{n}(z) e^{in\theta},
\end{equation}
where $J_n$ are $n^{th}$ order Bessel functions, to calculate the Shapiro steps and each step's width.  

We can integrate to solve for $\phi_A(\tau)$:
\begin{equation}
    \phi_A(\tau) = \phi_0 + \omega_0 \tau + z\sin(\omega \tau)
\end{equation}
where $\phi_0$ is an arbitrary integration constant, $z = 2eV_{1}/ \hbar \omega$, and $\omega_0 = 2 e V_{0}/\hbar$. 
Then we have
\begin{widetext}
\begin{align}
     \overline{I}_s&(\phi_{0}, \omega_0 = m\omega) = \sum_{n}(-1)^n [x_2\sin(\phi_0)J_n(z)\delta_{m,n} + x_4\sin(2\phi_0)J_n(2z)\delta_{2m,n} \nonumber \\ 
     &+ x_6\sin(3\phi_0)J_n(3z)\delta_{3m,n}  
     + x_8\sin(4\phi_0)J_n(4z)\delta_{4m,n} + x_1\sin(\phi_0/2)J_n(z/2)\delta_{m/2,n} \nonumber \\ 
     &+ x_3\sin(3\phi_0/2)J_n(3z/2)\delta_{3m/2,n}  
     + x_5\sin(5\phi_0/2)J_n(5z/2)\delta_{5m/2,n}+ y_2 \cos(\phi_0)J_n(z)\delta_{m,n} \nonumber \\ 
     &+ y_4\cos(2\phi_0)J_n(2z)\delta_{2m,n}  
     + y_6 \cos(3\phi_0)J_n(3z)\delta_{3m,n} + y_8\cos(4\phi_0)J_n(4z)\delta_{4m,n} \nonumber \\ 
     &+ y_1 \cos(\phi_0/2)J_n(z/2)\delta_{m/2,n} 
     + y_3 \cos(3\phi_0/2)J_n(3z/2)\delta_{3m/2,n} \nonumber \\
     &+ y_5 \cos(5\phi_0/2)J_n(5z/2)\delta_{5m/2,n}]
\end{align}
\end{widetext}

\section{Additional Data}

Figures~\ref{fig:S1} and \ref{fig:S2} show calculations of the average critical current and critical current difference corresponding to data in Fig.~1(d-h). Figures~\ref{fig:S3} and \ref{fig:S4} show the critical current difference corresponding to data in Fig.~3 and 4, respectively.

\bibliography{refs}   
\end{document}